\documentstyle[prl,aps,epsfig,multicol,amssymb,amsfonts]{revtex}
\tighten
\renewcommand{\narrowtext} 
{\begin{multicols}{2}\global\columnwidth20.5pc} 
\renewcommand{\widetext}
{\end{multicols}\global\columnwidth42.5pc} 
\newcommand{\be} {\begin{equation}}
\newcommand{\ee} {\end{equation}}

\begin{document} 
\draft 
\title{	Dimensionality dependence of the wave function statistics
	at the Anderson transition} 
\author{A.~Mildenberger$^1$, F.~Evers$^2$, and A.~D.~Mirlin$^{1,2,*}$ } 
\address{$^1$Institut f\"ur Theorie der Kondensierten Materie,
Universit\"at Karlsruhe, 76128 Karlsruhe, Germany}
\address{$^2$Institut
f\"ur Nanotechnologie, Forschungszentrum Karlsruhe, 76021 Karlsruhe,
Germany}
\date{\today}
\maketitle
\begin{abstract}
The statistics of critical wave functions at the Anderson transition in
three and four dimensions are studied numerically. The distribution of the
inverse participation ratios (IPR)  $P_q$ is shown to
acquire a scale-invariant form in the limit of large system size.
Multifractality spectra governing the scaling of the
ensemble-averaged IPRs are determined. Conjectures 
concerning the IPR statistics and the multifractality 
at the Anderson transition in a high spatial dimensionality are
formulated. 
\end{abstract}

\pacs{PACS numbers: 72.15.Rn, 71.30.+h, 05.45.Df, 05.40.-a} 
\narrowtext

A disordered electronic system in $d>2$ dimensions can be driven
from the phase of extended states to
that of localized states by
increasing the strength of disorder. This transition, bearing the name of
Anderson, is characterized by remarkably rich critical properties. In
particular, the eigenfunctions at the critical point show strong
fluctuations and represent multifractal distributions.

These fluctuations can be quantitatively characterized by a set of
inverse participation ratios (IPR), $P_q=\int d^dr\, |\psi({\bf r})|^{2q}$. 
Using the renormalization group in $d=2+\epsilon$ dimensions
($\epsilon\ll 1$), Wegner found \cite{wegner80}
that the ensemble-averaged IPRs, $\langle P_q \rangle$,
show at criticality an anomalous scaling with
respect to the system size $L$, 
$\langle P_q\rangle\propto L^{-\tilde{\tau}_q}$, where \cite{note0}
\be
\label{e1}
\tilde{\tau}_q = (q-1)d - q(q-1)\epsilon + O(\epsilon^4).
\ee
Equation (\ref{e1}) is written for the case of unbroken time-reversal
symmetry corresponding to the orthogonal ensemble which we consider
in the paper.
According to Eq.~(\ref{e1}), the fractal dimensions
$\tilde{D}_q=\tilde{\tau}_q/(q-1)$ are different from the spatial
dimension $d$ and depend on $q$, manifesting the multifractal
character of the wave function intensity
$|\psi({\bf r})|^2$ \cite{castpel}.
It is customary to characterize such a distribution by its singularity
spectrum determined by the Legendre transform of $\tilde{\tau}_q$,
yielding \cite{castpel,wegner87}
\be
\label{e2}
\tilde{f}(\alpha) = d - {(d+\epsilon-\alpha)^2/4\epsilon} +
O(\epsilon^4)\ .
\ee

In the present paper we  study how the wave function
statistics and, in particular, the multifractality spectrum evolve
with increasing spatial dimensionality, when the transition shifts
from the weak-disorder range (as in $d=2+\epsilon$) to the
strong-disorder one.
A numerical study of the Anderson transition in higher-dimensional
systems is of special interest since no analytical results for the
critical behavior in high $d$ are available. In contrast to
conventional second-order phase transitions for which the mean-field
treatment becomes valid above the upper critical dimension $d_c$, so
that the critical exponents are $d$-independent for $d\ge d_c$,
the usual mean-field approach fails in this case. The model has been
solved on the Bethe lattice \cite{bethe} which is
believed to correspond to the limit $d=\infty$. While for the
critical exponent of the localization length the conventional
mean-field value $\nu=1/2$ was obtained, a very unusual non-power-like
critical behavior of other quantities was found (namely, an
exponential vanishing of the diffusion constant and a jump in the
inverse participation ratios $\langle P_q\rangle$ with $q=2,3,\ldots$
at the mobility edge). 
It was, however, shown in \cite{mf94}
that these results are intimately related to the spatial structure of
the Bethe lattice. On this basis, it was argued that although the
symmetry-breaking description of the transition in terms of an
order-parameter function \cite{bethe}
is of general validity, the peculiar critical
behavior found in \cite{bethe} is an artifact of the Bethe lattice and
should take a power-law form for any finite $d<\infty$. 
Assuming that the critical indices should match at $d\to\infty$ the
Bethe lattice behavior (which corresponds to $\tilde{\tau}_q=0$ for
$q\ge 2$ and
to the value $s=\infty$ for the conductivity exponent), one then
concludes that the upper critical dimension is $d_c=\infty$
\cite{mf94} (see also \cite{cast86}).  
While available numerical results
for the value of $\nu$ \cite{schreiber96,zhar98} and for the form of the
critical level statistics \cite{zhar98} in four dimensions (4D)
are consistent with this
conjecture, its rigorous justification and a
systematic  analytical study of the transition for large $d$ are
still missing. 

To investigate the dimensionality dependence of the critical
statistics of wave functions, we have
performed numerical simulations of 3D and
4D tight-binding models with periodic
boundary conditions and a box distribution of site energies. 
The critical values
of the disorder are known to be 
$W_c \simeq 16.5$ for 3D \cite{slevin00} and $W_c\simeq 35$ for
4D \cite{schreiber96,zhar98,markos94}.
We have calculated wavefunctions with energy close
to zero by diagonalizing the Hamiltonian using efficient
numerical packages \cite{num3,num1}. Thereby
we could average over an ensemble that contained
typically $10^3$ samples. From every sample
128 wavefunctions with energy close to zero have been 
taken into account. 

\begin{figure}
\includegraphics[width=0.9\columnwidth,clip]{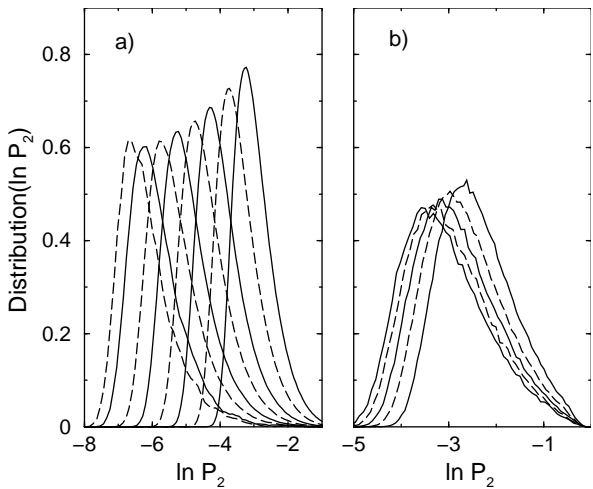}
\caption{IPR distribution in (a) 3D (system sizes 
         $L=8,$\newline$11,16,22,32,44,64,80$) and (b) 4D ($L=8,10,12,14,16$).}
\label{fig1} 
\end{figure}

\begin{figure}
\includegraphics[width=0.8\columnwidth,clip]{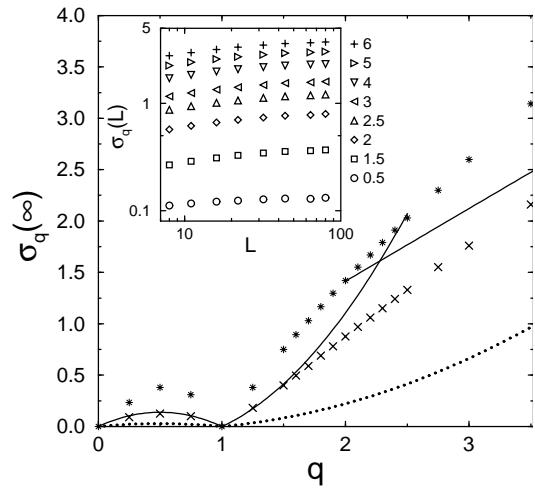}
\caption{The rms deviation $\sigma_q$ of $\ln P_q$ extrapolated to
$L\to\infty$ in 3D ($\times$) and 4D (stars). The dotted line is the
analytical result (\ref{e2a}) for $\epsilon=0.2$; the full lines
represent Eqs.~(\ref{e2a}) and (\ref{e2c}) with $\epsilon=1$. 
Inset: evolution of $\sigma_q$ with $L$ in 3D. The leading finite
size correction of all data has the form
$L^{-y}$ with $y=0.25\div0.5$ for 3D and $y=0.1\div0.4$ in 4D.
The numerical error in the extrapolated values of $\sigma_{q}(\infty)$
is as big as 10\% due to the uncertainty in $y$. }
\label{fig2} 
\end{figure}

We begin by showing in Fig.~1 the evolution of the distribution 
${\cal P}(\ln P_2)$ in 3D with the linear size $L$ of the system from
$L=8$ to $L=80$. It is clearly seen that at large $L$ the distribution
acquires a scale-independent limiting form and simply shifts along the
$x$-axis without changing its shape. The scale-invariance of the IPR
distribution at criticality was conjectured in \cite{fm95a}. This
conjecture was questioned in \cite{parshin99}, where numerical
simulations of the IPR distribution in 3D were performed, with the
conclusion that the fractal dimension $D_2$ is not a well defined
quantity but rather shows strong fluctuations with
root mean square deviation
${\rm rms}(D_2) \sim 1$. In terms of the IPR distribution the
statement of \cite{parshin99} would mean that the rms
deviation $\sigma_2 \equiv [{\rm var}(\ln P_2)]^{1/2}$
scales in the same way as $\langle-\ln P_2\rangle\propto \ln L$ in the
limit of large $L$. However, a detailed analytical and numerical study
of the wave function statistics in a family of critical power-law
random banded matrix (PRBM) ensembles has shown \cite{prbm} that the
IPR distribution ${\cal P}(\ln P_q)$ is scale-invariant, corroborating
the conjecture of \cite{fm95a}. It was argued that this is a general
feature of the Anderson transition, and the conclusion of
Ref.~\cite{parshin99} was criticized as based on a not sufficiently
careful numerical analysis of data obtained for too small systems.
This expectation was supported  by a very recent numerical study of
the 3D Anderson transition \cite{cuevas} where  a trend to
saturation of the width of the distribution ${\cal P}(\ln P_2)$ with
increasing length $L$ was found. 
Our data (Fig.~\ref{fig1}a) demonstrate the
scale-invariance of the limiting distribution even more convincingly,
since we have reached considerably larger values of $L$. The same
conclusion can be drawn from the inset of Fig.~\ref{fig2} where the
saturation of the rms deviation $\sigma_q(L)$ of the distribution 
${\cal P}(\ln P_q)$ at $L\to\infty$ is demonstrated. Qualitatively
similar results are obtained for the IPR distribution in 4D
(Fig.~\ref{fig1}b). Though a tendency to the 
saturation with increasing $L$ is
clear in this case as well, the full saturation has not been reached,
in view of smaller linear sizes of the system as compared to 3D.
 
To visualize the strength of the IPR fluctuations, we show
in the main panel of Fig.~\ref{fig2}  the values of the rms
deviations $\sigma_q$, which characterize the width of the
distribution functions ${\cal P}(\ln P_q)$, extrapolated to
$L\to\infty$. In $d=2+\epsilon$ dimensions $\sigma_q$ can be
calculated analytically following \cite{fm95a}, with the result
\be
\label{e2a}
\sigma^2_q = 8\pi^2 a_2 \epsilon^2 q^2(q-1)^2, 
\qquad q\ll q_c =(2/\epsilon)^{1/2}, 
\ee
where $a_2$ = 0.00387 for the periodic boundary conditions.
Here $q_c$ is the value of $q$ corresponding to the root $\alpha_-$ of
the singularity spectrum $\tilde{f}(\alpha)$, {\it i.e.} 
$q_c = \tilde{f}'(\alpha_-)$ and $\tilde{f}(\alpha_-)=0$. 
For $q\gg q_c$ the IPR distribution is dominated by its slowly
decaying power-law ``tail'' 
\be
\label{e2b}
{\cal P}(P_q/P_q^{\rm typ})\propto 
(P_q/P_q^{\rm typ})^{-1-x_q}, \qquad P_q\gtrsim P_q^{\rm typ},
\ee
where $x_q =q_c/q$ \cite{prbm}.
This yields $\sigma_q = q/q_c$, or in $2+\epsilon$ dimensions,
\be
\label{e2c}
\sigma_q = (\epsilon/2)^{1/2} q, \qquad q \gg (2/\epsilon)^{1/2}.
\ee
In addition to the numerical results for 3D and 4D, we present 
in Fig.~\ref{fig2} the result (\ref{e2a}) of the $\epsilon$-expansion
for $d=2.2$. Furthermore, we show the small-$q$ and large-$q$
analytical asymptotics (\ref{e2a}) and (\ref{e2c}) with $\epsilon=1$. 
Of course, the $\epsilon$-expansion is only justified parametrically
for $\epsilon\ll 1$. Nevertheless, we see that it still describes very
reasonably the IPR fluctuations at the Anderson transition in 3D. 

Having demonstrated that the distributions of the IPRs $P_q$ are
scale-invariant at criticality, so that the fractal dimensions are
well defined, we are prepared to analyze the form of the multifractal
spectra. We
use the numerical procedure described in Ref.~\cite{qhe}, where it was
applied to the quantum Hall plateau transition. Specifically, 
we calculate the ensemble-averaged IPR $\langle P_q\rangle$ for
different system sizes $L$, extract $\tilde{\tau}_q$ and
$\tilde{f}(\alpha)$ and extrapolate to the thermodynamic limit
$L\to\infty$ in order to eliminate the finite-size corrections.
We refer the reader to Ref.~\cite{qhe} for more details of the
procedure and for a discussion of its advantages as compared to the
box-counting calculations performed in earlier publications.

\begin{figure}
\includegraphics[width=0.8\columnwidth,clip]{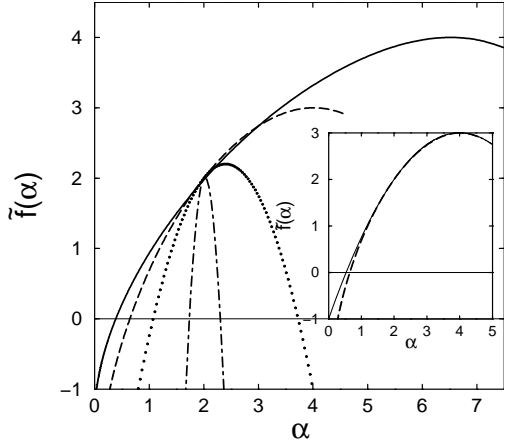}
\vspace{3mm}
\caption{Singularity spectrum $\tilde{f}(\alpha)$ in 3D (dashed) and
4D (full line). To illustrate the evolution of the spectrum from $d=2$ to $d=4$,
analytical results for $d=2+\epsilon$ are shown for $\epsilon=0.2$
(dotted) and $\epsilon=0.01$ (dot-dashed). Inset: comparison between
$\tilde{f}(\alpha)$ for 3D and
Eq. (\ref{e2}) with $\epsilon=1$ (solid).}
\label{fig3} 
\end{figure}
\begin{figure}
\includegraphics[width=0.7\columnwidth,clip]{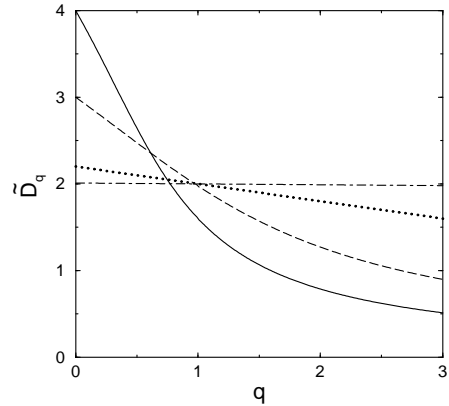}
\caption{Fractal dimensions $\tilde{D}_q$ in 3D (dashed) and 4D (full
line). Analytical 
results for $d=2+\epsilon$ with $\epsilon=0.2$ (dotted) and
$\epsilon=0.01$ (dot-dashed) are also shown. }
\label{fig4} 
\end{figure}

The obtained results for $\tilde{f}(\alpha)$ in 3D and 4D are shown in
Fig.~\ref{fig3}. To illustrate the evolution of the spectrum in the whole
range from $d=2$ to $d=4$, the analytical results (\ref{e2}) for
$2+\epsilon$ dimensions with $\epsilon=0.2$ and
$\epsilon=0.01$ are also shown. 
In Fig.~\ref{fig4} the corresponding
results for the fractal dimensions $\tilde{D}(q)$ are presented.
We see that with increasing dimensionality the singularity
spectrum $\tilde{f}(\alpha)$ broadens. This is not surprising: with
increasing $d$ the transition moves further in the region of strong disorder,
implying stronger multifractality. What is much less
obvious is that in the range $\alpha \lesssim 2$ the $\tilde{f}(\alpha)$ curve 
shifts to the left with increasing $d$. This corresponds to the fact
that the fractal exponents $\tilde{D}_q$ with $q\gtrsim 1$ 
decrease with increasing $d$. In particular, for the exponent $\tilde{D}_2$
determining the spatial dispersion of the diffusion coefficient at
criticality we find $\tilde{D}_2=1.3\pm0.05$ in 3D and $\tilde{D}_2=0.9\pm0.15$ in
4D \cite{note1}. It is worth mentioning that the $\epsilon$-expansion
(\ref{e2}) with $\epsilon=1$ describes the 3D spectrum remarkably
well (though with detectable deviations, see inset of
Fig. \ref{fig3}). In particular, the position of the maximum,
$\alpha_0=4.03\pm 0.05$ is very close to its value $\alpha_0=d+\epsilon$
implied by (\ref{e2}).
As expected, in 4D the deviations from the parabolic shape are much
more pronounced and $\alpha_0=6.5\pm0.2$ differs noticeably from $6$.

The obtained value of $\tilde{D}_2=1.3\pm0.05$ in 3D is considerably smaller
than what was found in the earlier numerical studies \cite{d2}
where the values
in the range $1.4\div 1.8$ were reported. 
The reasons for this are as follows. In the earlier works the spectrum
$f(\alpha)$ of individual eigenstates was studied via the box-counting
procedure. While in the limit $L\to\infty$ the spectrum $f(\alpha)$
defined in this way should reproduce the part of the
$\tilde{f}(\alpha)$ curve lying above the $x$-axis,  for a
finite $L$ the finite-size effects affect 
$f(\alpha)$ strongly, especially in the region close to the zero
$\alpha_-$ of $\tilde f(\alpha)$.
As a result, such a method not only fails to yield the exponents
$\tilde{D}_q$ with $q>q_c$ but also
leads to very large errors in determination of $D_q$ with $q$ smaller
than but close to $q_c$. In particular, in 3D and 4D we find that
$q_c$ is close to 2, explaining large errors in earlier results for
$q=2$ in 3D. As to the 4D case, we are not aware of any previous
studies of the wave function statistics at the Anderson transition.  

Let us discuss now what our findings imply for the high-$d$ behavior
of the critical wave function statistics. First of all, we notice that
our numerical observation that $\tilde{D}_q$ with $q\gtrsim 1$ decrease with
increasing dimensionality confirms the above-mentioned expectation
that they should
tend to zero in the limit $d\to\infty$. In view of this, an analogy
with the PRBM ensemble with the parameter $b\ll 1$
is very instructive. This ensemble
can be considered as describing a 1D chain with random
long-range hopping whose rms amplitude is $b/r$ where $r$ is the
distance. The model is critical for arbitrary $0<b<\infty$ and can
be studied analytically in both limits $b\gg 1$ and $b\ll 1$. The
latter case is relevant to the issue under discussion, and we remind
the reader of the key results
\cite{prbm} on  the wave function statistics at $b\ll 1$.
(The case $b\gg 1$ is to a large extent analogous to the Anderson
transition in $2+\epsilon$ dimensions with $\epsilon\ll 1$.) 
Specifically, the scale-invariant critical distribution 
${\cal P}(P_q/P_q^{\rm typ})$ becomes $b$-independent in the limit
$b\ll 1$. Furthermore, the exponents $\tilde{\tau}_q$ with $q>1/2$ 
are proportional to $b$ in the small-$b$ limit,
\be
\label{e3}
\tilde{\tau}_q=2b\tilde{T}_q, \qquad
\tilde{T}_q = 2{\Gamma(q-1/2)/\pi^{1/2}\ \!\Gamma(q-1)}.
\ee
Correspondingly, the singularity spectrum $\tilde{f}(\alpha)$ acquires
for $\alpha\ll 1$ the form 
\be
\label{e4}
\tilde{f}(\alpha) = 2b \tilde{F}(\alpha/2b)\ ,
\ee
where $\tilde{F}(A)$ is the Legendre transform of $\tilde{T}(q)$,
with the asymptotics $F(A)\simeq -1/\pi A$ as $A\to 0$ and
$F(A)\simeq A/2$ as $A\to\infty$. 
The smallness of the fractal exponents,  $\tilde{\tau}_q\ll 1$, reflects a
very sparse structure of the eigenstates formed by resonances (spikes) with a
hierarchy of distances between them, $r_1\ll r_2\ll \ldots$,
such that $\ln r_{i+1}/r_i\sim 1/b$ \cite{prbm}. 

After this reminder we return to the Anderson transition in high
dimensionality. The smallness of the fractal exponents,
$\tilde{D}_q\ll 1$  at $d\gg 1$ implies, in close similarity to
the PRBM model, that the corresponding
critical eigenstates have the resonance structure with a
hierarchy of scales  $\ln r_{i+1}/r_i\sim 1/b_d$ and with a
dimensionality-dependent parameter $b_d$ satisfying $b_d\to 0$ as
$d\to\infty$. 
%
%
(In contrast to the PRBM model, where the emergence of resonances was
due to direct long-range hopping processes, now it should be
determined by interference of all possible paths connecting two sites
on a $d$-dimensional lattice.) 
%
The sparse resonance structure of 
eigenstates allows us to use the analogy with the PRBM model and to 
make the following conjectures:
(i) The scale-invariant distribution ${\cal P}(P_q/P_q^{\rm
typ})$ for $q>1/2$ becomes essentially $d$-independent at $d\gg 1$.
(ii) The fractal exponents take at $d\gg 1$ and for $q>1/2$ the form
$\tilde{\tau}_q(d)=2b_d\tilde{T}_q$, {\it i.e.} they depend on $d$
only through the overall factor $b_d$ (satisfying $b_d\to 0$ as
$d\to\infty$).
Correspondingly, the singularity spectrum
scales at large $d$ as $\tilde{f}(\alpha) = 2b_d
\tilde{F}(\alpha/2b_d)$. 
%
%
Though we are fully aware of a speculative character of these
conjectures at the
present stage, we expect that they will stimulate further
numerical and analytical work, which should finally resolve the challenge
posed by the Anderson transition in high $d$.

In conclusion, we have studied the statistics of critical wave functions
at the Anderson transition in 3D and 4D. The distribution of the
inverse participation ratio $P_q$ was demonstrated to
acquire a scale-invariant form in the limit of large system size.
As a convenient measure of the strength of the IPR fluctuations, we have
evaluated the rms deviation $\sigma_q$ of $\ln P_q$ and found a result
matching well analytical predictions for $2+\epsilon$ dimensions with
$\epsilon\ll 1$. 
Calculating the ensemble-averaged IPR values, $\langle P_q\rangle$,
we have determined the spectrum of fractal dimensions
$\tilde{\tau}_q\equiv \tilde{D}_q(q-1)$ and the singularity spectrum
$\tilde{f}(\alpha)$ characterizing the multifractal 
properties of the wave functions. In particular, we have found
$\tilde{D}_2=1.3\pm0.05$ in 3D and $\tilde{D}_2 =0.9\pm0.15$ in
4D. More generally, 
our results indicate that the dimensions $\tilde{D}_q$ with 
$q \gtrsim 1$ decrease with increasing spatial dimensionality $d$.
On this basis, we
have formulated two conjectures concerning the wave function
statistics at criticality in the large-$d$ limit.

It is a pleasure to acknowledge help with the implementation
and usage of the {\it Watson sparse matrix package} by
A. Gupta and M. Krauss. 
This work was supported by the SFB195 and the Schwerpunktprogramm
``Quanten-Hall-Systeme'' der Deutschen Forschungsgemeinschaft.

\end{multicols}

\begin{references}
\bibitem{note0}
Following the notations introduced in
\cite{prbm}, we indicate by tilde that the index $\tilde{\tau}_q$
in (\ref{e1}) corresponds to the ensemble-averaged IPR
rather than to the typical one, $P_q^{\rm typ}$. 
For a detailed discussion of the necessity to
distinguish between $\langle P_q\rangle$ and 
$P_q^{\rm typ}$ the reader is referred to Ref.~\cite{prbm}.
\end{references}
\end{document}